\documentstyle[preprint,prd,aps,floats,epsfig]{revtex}
\tighten
\begin{document} 
\input psfig
\draft
\preprint{
\begin{tabular}{rr}
CfPA/97-th-10\\
IC-96-97/23\\
Submitted to {\em PRD}
\end{tabular}
}

\title{Cumulants as non Gaussian qualifiers}
\author{Pedro G. Ferreira$^{1}$, Joao Magueijo$^2$ and Joseph Silk$^{1,3,4}$}
\address{$^{(1)}$Center for Particle Astrophysics, Department of
Astronomy and Physics,
University of California, Berkeley  CA 94720-7304,USA\\
$^{(2)}$Theoretical Physics, The Blackett Lab.,
Imperial College, Prince Consort Road, London, SW7 2BZ, U.K.\\
$^{(3)}$Insitut d'Astrophysique, 98 bis, Boulevard Arago, Paris, France\\
$^{(4)}$Insitute of Astronomy, Madingley Road, Cambridge CB3 0HA, U.K.
}
\maketitle
\begin{abstract}
\narrowtext
We discuss the requirements of good statistics for quantifying
non-Gaussianity in the Cosmic Microwave Background. 
The importance of rotational invariance
and statistical independence is stressed, but we show that these are sometimes
incompatible. It is shown that the first of these requirements
prefers a real space (or wavelet) formulation, whereas the latter
favours quantities defined in Fourier space. Bearing this
in mind we decide to be eclectic and define two
new sets of statistics to quantify the level of
non-Gaussianity. 
Both sets make use of the concept of cumulants of a distribution.
However, one set is defined in real space, with reference to the wavelet
transform, whereas the other is defined in Fourier space.
We derive a series of properties concerning these statistics for a Gaussian
random field and show how one can relate these quantities to
the higher order moments of temperature maps. 
Although our frameworks lead to an infinite hierarchy of quantities we show 
how cosmic variance and experimental constraints give a natural truncation
of this hierarchy. We then focus on the real space statistics and 
analyse the non-Gaussian signal generated by points sources obscured
by large scale Gaussian fluctuations. We conclude by discussing the
practical implementations of these techniques.
\end{abstract}

\date{\today}


\renewcommand{\thefootnote}{\arabic{footnote}}
\setcounter{footnote}{0}
\section{Introduction}
One of the primary goals of cosmology is a complete characterization
of the seeds that led to the formation of structure. From an accurate
understanding of the statistics  of fluctuations we may be able
to glean information about the physical origin of these seeds, their
evolution and even find a precise measure of the parameters
that characterize the space-time on which they live. A laboratory
from which we can obtain detailed estimates is supplied by the cosmic
microwave background (CMB). Much effort has
focused on optimal estimates of the fluctuations' power spectrum 
given the harsh
realities of experimental data; one is faced with foreground
contamination, incomplete sky coverage, and instrumental noise, which have
to be incorporated into data analysis. A simplifying assumption has
been that fluctuations in the CMB are Gaussian, allowing the
development of sophisticated techniques 
for estimating cosmological parameters.

With current developments in experimental CMB physics, we will now
be in a position to analyse very large data sets, with information
about large patches of the sky measured with very high resolution
and sensitivity. This means that we are in a position to seriously
test some of the assumptions that have gone into the techniques
that have been developed, in particular, whether the sky is
really Gaussian. If the sky is Gaussian indeed, then current techniques 
for estimating the power spectrum will become watertight methods. Should
deviations from non-Gaussianity be detected, clearly one should
start again. Furthermore the power spectrum would then not be the 
end of the story in the quest for a statistical characterization
of the fluctuations.

In the past this task has been tackled in a variety of ways
subject to very different philosophies. One  approach has been to choose
a statistic which is easy to describe for a Gaussian random field and then
try to quantify, by means of this statistic, what are the chances 
that a given data set comes from an underlying Gaussian ensemble. The
well known examples are 
peaks' statistics \cite{BE,Jusk}, topological tests \cite{Pcoles,gott}, 
the 3-point correlation function \cite{kogut} and 
skewness and kurtosis \cite{scaramelo}. Another approach has been
to devise statistics which are good discriminators between Gaussian
skies and specific non-Gaussian rivals. Such is the case in much
of the techniques involved in looking for topological defects \cite{VS}, such
as strings and textures or even foregrounds, such as point sources.

These approaches have their merits. Non-Gaussian tests are very easy
to implement even in the context of very large data sets. Also
reducing the whole issue
to a single statistic allows one to concentrate on devising the 
statistic ideally suited for detecting a given, pre-known, type
of non-Gaussianity. This is somewhat reminiscent of pattern recognition:
if we already know what we are looking for, we may improve our
chance of detecting an existing predefined pattern inside a noisy data-set.
 
One can, however, take a more humble approach, which is to admit that
we have little idea of what the underlying probability distribution
function of the CMB is. It then becomes necessary to devise as
complete a framework as possible, without prejudices with regards
to testing rival models, or ease of computation with regards to testing
non-Gaussianity. This is an alternative approach, which supports as
its underlying  philosophy the quest for ruling out or detecting ${\it generic}$ 
non-Gaussianity.

The most well established formalism following 
this alternative philosophy is the
{\it n-point formalism}. We briefly describe this formalism.
Consider the CMB anisotropies, $\frac{\Delta T}{T}({\bf x})$ to be 
statistically isotropic random field, defined on the sky. 
We consider CMB data in the small angle limit, when projecting onto
a planar patch is suitable. Since data may come in either real or
Fourier space we want to address the problem of non-Gaussianity
in terms of both representations. 
We shall use the convention:
\begin{equation}\label{fourier}
  \frac{\Delta T({\bf x})}{T}
={\int {d{\bf k}\over 2\pi}a({\bf k})e^{i{\bf k}\cdot{\bf x}}}
\end{equation} 
The $n$-point correlation function is defined as the expectation
value of the product of any $n$ temperatures. Translational and
rotational invariance make redundant the position of one of the points
and the direction of another.
Hence the $n$-point function may be written as a function of
$(x_2, {\bf x}_3,\cdots,{\bf x}_n)$ in the form
\begin{equation}
  C^n(x_2, {\bf x}_3,\cdots,{\bf x}_n)={\langle 
  \frac{\Delta T({\bf x}_1)}{T}...\frac{\Delta T({\bf x}_n)}{T} 
  \rangle} \label{eq:npoint}
\end{equation}
The 2-point correlation function and its Fourier transform, the 
angular power spectrum $C(k)$,  are well-known.
The angular power spectrum may be generalized for $n>2$ by
Fourier analysing the $n$-point  function 
\begin{eqnarray}\label{cn1}
  &C^n&(x_2, {\bf x}_3,\cdots,{\bf x}_n)= \nonumber \\
&&\int \frac{dk_2}{(2\pi)^{1/2}}
  \cdots \frac{d{\bf k}_n}{2\pi}C^n(k_2, \cdots,{\bf k}_n)
  {e^{ik_2 x_2} \cdots e^{i{\bf k}_n\cdot{\bf x}_n}}
\end{eqnarray}
In spite of all its success it is argued in \cite{fermag}
that this framework is not systematic and is plagued by redundancy. 
In principle, one can calculate an infinite
number of n-point functions, and there is no criteria where to
truncate such an evaluation. If one has a finite data set, then
many of these quantities will be algebraically dependent on each other. 
There is a practical additional problem: to estimate the m-point correlation
function, one needs ${\cal O}(N_{pix}^m)$ operations, clearly a 
large number for the expected large data-sets.

A possible remedy to these problems was outlined in \cite{fermag}.
Here we follow up on this work, but along a rather different angle.
We wish to set up a practical non-redundant formalism for encoding generic
non-Gaussianity, but to start with all we define are the requirements
an ideal such formalism would satisfy. We then look at the formats
in which data comes, and within the available descriptions we try to accommodate
our requirements the best we can. The outcome is necessarily an
eclectic mixture of techniques. These, we hope, will be practical
devices subject to as little prejudice as possible. 

\section{Requirements for the ideal statistics} 
Five major requirements will ensure that the chosen quantities afford a
sensible statistical description of the random field. Depending on 
the particular data set available, and on the particular theory one beloves,
one may give more or less emphasis to each of these requirements.

\subsection{Invariance}
An essential requirement is that any statistic one defines is
rotationally and translationally invariant. If we define our data set to be
a set of pixels $\{\Delta T({\bf x})\}$ and the statistic $\cal G$
to be a functional then
\begin{eqnarray}
{\cal G}[\Delta T(R{\bf x}+{\bf a})]={\cal G}[\Delta T({\bf x})]\nonumber
\end{eqnarray}
The reason underlying this criterion is the assumption that the CMB can
be described as a statistically isotropic random field. The ensemble
of all data sets $\{\Delta T({\bf x})\}$
remains the same under rotations and translations. 
Therefore it is convenient to define  statistics which,
when applied on a given realization, do not change if 
rotations and translations are performed upon this realization. 
By doing so we are probing more realizations in the
ensemble, since the ensemble, being isotropic, replicates each realization
into all the realizations related by the symmetry transformation.
A good example of this requirement being enforced is the angular
power spectrum $C_\ell$ which is independent of the way in which the
axes are defined.

This requirement is essential
in the standard Big Bang cosmology and within all-sky experiments,
but there may be grounds for dropping it otherwise. 
If we look at small patches of the sky, then the existence
of an observation window is already breaking
the translational and rotational invariance. Also a few examples of
fundamentally anisotropic fluctuations have been discussed in the 
past \cite{barrow,ped,fermag2}.

In the light of this argument we shall enforce this requirement in 
the construction in Section IV but not in the construction in Section V.

\subsection{Statistical independence for a Gaussian process}
Another requirement is that any set of statistics that
one defines is made up of quantities which, for a Gaussian theory,
are statistically independent, and that one of these quantities be the
power spectrum (which completely describes a Gaussian theory). 
In fact if some of the quantities we define are dependent
then we are over counting degrees of freedom along which the 
theory is allowed to be non-Gaussian.

Such a framework was proposed in \cite{fermag}, where one provides
a transformation from the Fourier space values of the temperature
anisotropies into a complete set of independent quantities. 
This has the advantage of identifying the correct
amount of information that one can correctly assess from a 
finite data set. 

This requirement may however be satisfied in practice, for
large data sets, even if not in theory by the formalism.
A large data set will be an assumption we shall make in this paper
in order to satisfy this criterion. 

\subsection{Scale dependence}
It has become clear that different physical processes are
important on different physical scales. These different
processes may have very different statistical properties.
For example, the  distribution of point sources will be
a Poisson process, while for high wavenumbers
the surface of last scattering will have an exponentially
suppressed power spectrum. If one were to look at the
sky at sub arc minute scales, any given pixel would be the
sum of these two Gaussian and non-Gaussian distributions.
It is often the case that the Gaussian component
dominates the non-Gaussian one on some scales but not on others.
It is therefore desirable to define
statistics which are scale dependent. In the application given in
the Section VI we show how this feature may be decisive in the
ability of any statistic to pick subtle non-Gaussian features.

\subsection{Additivity}
It may be useful if the statistics we define are additive,
in the sense that if $\Delta T_1({\bf x})$ and $\Delta T_2({\bf x})$
are maps coming from two independent random processes then
the statistic $\cal G$ will satisfy
\begin{eqnarray}
{\cal G}[\Delta T_1({\bf x})+\Delta T_2({\bf x})]=
{\cal G}[\Delta T_1({\bf x})]+{\cal G}[\Delta T_2({\bf x})]\nonumber
\end{eqnarray}
Lack of additivity is 
one of the shortcomings of the formalism in \cite{fermag}.

Additivity may be useful firstly because one may sometimes
want to combine information on different scales. For instance
a given effect may be present in a band of scales, within
which the power spectrum may vary, rather than in just a single scale.
For instance in \cite{fermag} one finds a set of transverse spectra which in
the language of $a^\ell_m$ complement the power spectrum $C_\ell$
(which tells us how much power there is on the scale $\ell$)
with a transverse spectrum $B^\ell_m$, which tells us how the power
on the scale $\ell$ is distributed in angle $m$. Such an approach
has the problem that few modes may then contribute to the $B^\ell_m$.
Better still would be to find a truly transverse spectrum $B_m$
which would average over a certain range of scales $\ell$ for
a fixed direction $m$. Such a $B_m$ spectrum would be truly 
orthogonal to the power spectrum $C_\ell$ description. Such
construction cannot be provided by \cite{fermag} because the
quantities defined there are not additive. For additive
statistics, however, extending non-Gaussian spectra over scales
to non-Gaussian spectra over bands of scales
is a trivial operation. Such a construction is described in 
Section~\ref{fouriers}.

Another motivation for additivity comes from networks of
non-Gaussian structures which are a Poisson process of individual
non-Gaussian structures. Such networks are often globally complicated
but their individual components are simple. For instance a cosmic string network is
a bit of a mess globally, but is made up of essentially simple
elements, say segments of Brownian strings. It is for this reason that the
formalism in \cite{fermag} is really better suited for small
fields, where the number of non-Gaussian objects is never
larger than one. For a large field \cite{fermag} provides
a rather complicated description. Again an additive statistic
would not encounter this problem. If the individual object
has a simple description within the formalism, then the same
would be true for a network of such objects. Another way to phrase
this concern is to say that it is useful to define statistics
sensitive to local rather than global features. 

We will show in Section VI this criterion at work in the context of point
source subtraction. Point sources in large fields are globally
complicated but locally extremely simple. We shall enforce this 
criterion in this context by combining two tools. Firstly we shall
make use of additive cumulants. Secondly we shall introduce the concept
of scale by means of the local wavelet transform.


\subsection{Computational efficiency}
There are practical considerations. As mentioned above, estimating
higher order statistics within the n-point formalism is very demanding
on computation capabilities. One needs efficient methods which
will be manageable with future million pixel data sets and 
available computer resources. 

Related to this issue is the quest for comprehensive but non-redundant 
statistics. This is sometimes a problem. One can show that a distribution
may be Gaussian up to a very large moment and then be non-Gaussian
(although the opposite is not possible, see \cite{ko}). An infinite
and largely repetitive series of moments is therefore required for
completeness. We will see however how cosmic variance provides a natural
cut-off for what in principle is a infinite series of statistics. 
The idea is that if cosmic variance goes above a given level there is
no practical way in which we could detect non-Gaussianity, given the
fact that we only have one sky.

\subsection{Overall picture}
As one would expect, it is difficult to reconcile all these
requirements. Some of them are even incompatible. For example,
statistical independence leads one to work in Fourier space,
where statistical isotropy and homogeneity enforce statistical
independence of the different modes. On the other hand, 
Fourier space is a very non-local transformation, and so any 
statistic defined in Fourier space will be sensitive to the 
global properties of the sample. As stated before this may entail 
the awkward recognition of globally complicated networks of
essentially simple components. Enforcing
translational invariance, while keeping additivity, is also
impossible to do in Fourier space.

We will start from a simple idea: to refine the notion of
one point distribution function of $\frac{\Delta T}{ T}({\bf x})$
in such a way as to incorporate as many of these requirements as possible. 
The fundamental idea is to calculate cumulants, or combinations
of cumulants defined on various transforms on the data sets.
Depending on the priorities, these transforms will be in real
space or Fourier space. There are different ways in which this simple 
idea may be implemented. Different alternative will lead to
favouring some of the above properties 
over others. It will be instructive to consider a few alternatives.


\section{Histograms and Cumulants}\label{cuhi}
We will take as our starting point the one-point distribution
function of $\frac{\Delta T}{ T}({\bf x})$. If we were to consider an
ensemble of realizations, we would be able to characterize this
distribution completely. By inspecting the $\frac{\Delta T}{T}({\bf x})$
histogram of realizations one can then see if the distribution 
function is Gaussian or not. However, the histogram is but a graphical device.
The algebraic statement corresponding to this non-Gaussianity test 
consists of studying the cumulants.

The cumulants of a sample (or of a distribution) are first introduced
in an attempt to achieve additivity.
We can define the $r^{th}$ moment of the $\frac{\Delta T}{T}({\bf x})$
distribution to be
\begin{eqnarray}
\mu_r(\frac{\Delta T}{ T})=\langle(\frac{\Delta T}{ T})^r\rangle
\nonumber
\end{eqnarray}
and from Eq. (\ref{eq:npoint}) we see that $\mu_r=C^r(0,\cdots,0)$.
The moments, if they exist, fully quantify the distribution as they appear as
coefficients in the Taylor expansion of the characteristic
function $\phi(t)$ (the characteristic is essentially the Fourier transform
of the distribution function). The characteristic 
satisfies $\phi_{A+B}=\phi_A\phi_B$
and so the $\mu_r$ cannot be generally additive. The idea
behind the definition of cumulants consists of writing down polynomials
in the $\mu_r$ which are additive, $\chi_r(A+B)=\chi_r(A)+\chi_r(B)$.
The prescription is defined in \cite{ko} and consists simply
of taking the logarithm of the characteristic $\psi=\log\phi$.
Then $\psi_{A+B}=\psi_A+\psi_B$. If $\psi$ is expanded in power
series one obtains, as coefficients, a series of additive moments or cumulants
$\chi_r$. Moments and cumulants may be related by comparing the
expansions of $\psi$ and $\log\phi$. In general cumulants may 
be obtained from the moments using
\begin{equation}
  \chi_r=r!\sum_{k=1}^{r}(-1)^{k+1}{{1}\over k}\sum_{\nu_1=1}^{r-k+1}\cdots
  \sum_{\nu_k=1}^{r-k+1} {\mu_{\nu_1}
     \over\nu_1!}\cdots{\mu_{\nu_k}
     \over\nu_k!}
\label {eq:cumdef}
\end{equation}
where the indices $\nu_i$ must satisfy $\nu_1+\cdots+\nu_k=r$.
At this point one notices that we get more than what we bargained
for. For a Gaussian $\psi(t)=-\sigma^2t^2$ and so the cumulants
of a Gaussian must be zero for $r>2$. This is an added benefit
over the moments, which are not zero for a Gaussian for even orders,
but instead have the relatively complicated spectrum of values:
\begin{equation}
\mu_{2r}={(2r)!\mu_2^r\over 2^r r!}=(2r-1)!!\mu_2^r
=(2r-1)(2r-3)...5\times 3\mu_2^r
\end{equation}
Formula (\ref{eq:cumdef}) is not the best way to compute cumulants
from the moments. An efficient algorithm is given in \cite{ko},
where it is shown that
\begin{eqnarray}
\chi_r=(-1)^{r-1}\left| \begin{array} {cccccc} \mu_1 & 1& 0&0 & \cdots& 0\\
\mu_2 &\left(\begin{array} {c}1\\0\end{array} \right)\mu_1 &1&0&\cdots&0\\
\mu_3 &\left(\begin{array} {c}2\\0\end{array} \right)\mu_2 &\left(\begin{array} {c}2\\1 \end{array} 
\right)\mu_1& 1&\cdots&0\\
\cdots & \cdots &\cdots &\cdots&\cdots&1 \\
\mu_r & \left(\begin{array} {c}r-1\\0\end{array} \right) \mu_{r-1} 
& \left(\begin{array} {c}r-1\\1\end{array} \right) \mu_{r-2} &\cdots
&\cdots&
\left(\begin{array} {c}r-1\\r-2\end{array} \right) \mu_1 
\end{array}\right|\label{eq:det}
\end{eqnarray}
Dimensionless quantities may be constructed out of the cumulants
\begin{equation}
  {\bar \chi}_r={\chi_r\over \mu_2^{r/2}}
\end{equation}
of which the familiar ${\bar \chi}_3$ and ${\bar \chi}_4$ 
are known as the skewness and kurtosis.

The quantities $\chi_r$ and $\bar{\chi}_r$ have two useful properties,
regardless of the sample on which they are defined.
Firstly they are zero for a Gaussian probability distribution
function. If one considers the $\bar{\chi}_r$s then one can
quantify this  in a way which is independent of the power spectrum.
 Secondly the $\chi_r$s are are additive. This means that if
$\frac{\Delta T}{T}$ is the sum of many different processes, its
cumulants will be the sum of the cumulants of each process.
One can't have both of these properties, and depending on the
situation we will opt to work with $\chi_r$ or $\bar{\chi}_r$.

There are a few subtleties which should be considered when working
with these statistics and we will bear them in mind throughout the
paper. Firstly, care must be taken when estimating these quantities
from a limited sample. The simplest procedure is to define first estimators
for the moments
\begin{eqnarray}
{\hat \mu}_r=\frac{1}{N}\sum_i\left(\frac{\Delta T}{T}\right)^r_i \label{eq:mom}
\end{eqnarray}
where hat denotes an estimator and $N$ is the number of 
pixels used in the estimation. We can then 
use Eq. (\ref{eq:cumdef}) or (\ref{eq:det}) to define estimators for
the cumulants ${\hat \chi}_r$. 
It turns out that these estimators aren't centered,
i.e. $\langle{\hat \chi}_r\rangle\neq 0$ for a Gaussian distribution.
The value of the bias is a function of $N$ and 
in the case where one has large $N$
this is not a problem. On the other hand, 
for small samples, one can construct unbiased
centered estimators of the cumulants using what are known as
$\kappa$-statistics. The idea is to bypass estimating the moments and
define polynomials in the $\frac{\Delta T}{T}$
which do average to the cumulants. For the purpose of this paper we will
always consider the large N limit. In the  discussion we will consider
the limitations of such an assumption. 

There is an interesting, well known connection that can be made between
cumulants and the connected Greens functions of statistical physics. 
In the latter case one is interested in quantifying the
corrections that will be introduced if one modifies a Gaussian
theory (a theory where the action  can
be written as a Gaussian functional on the fundamental fields) by
introducing modifications. This can be done by looking at
the connected Greens functions of the theory, which are simply
the difference between each Green's function of the non-Gaussian
theory and the Green's function of that order if one assumes
the theory is Gaussian (using Wick's theorem). The cumulants will
be the zero lag values of these connected Green's functions.

Finally, there is one important point that must be addressed which
may be a shortcoming of any technique which uses cumulants as the
basis of non-Gaussianity. If we work out the covariance matrix
for cumulants we find that it is not strictly diagonal. However we
find that
\begin{eqnarray}
\langle {\hat \chi}_r {\hat \chi}_{r'}\rangle =
\frac{r!}{N}\sigma^{2r}
\delta_{rr'}+{\cal O}(\frac{1}{N^2})\label{covcu}
\end{eqnarray}
The structure of the off-diagonal terms is simple to understand,
in light of Wick's theorem: if $r+r'$ is odd, covariance is strictly
zero while if $r+r'$ is even, it is proportional to $1/N^2$. This
means that in the limit of large $N$ (the realm of large data sets we
are considering in this paper) the set of cumulants are a
set of independent quantifiers of non-Gaussianity. The structure of
the covariance matrix also gives us a prescription at which we
can truncate, in $r$ the set of cumulants we should calculate
for a given data set. By defining a maximum variance allowed
we constrain   $r!/N$ to be less than some value, thereby truncating
the cumulants series at some value $r$.

\section{Real Space Statistics}\label{real}
Having defined cumulants we now address the issue of the sample
on which they should be computed. This depends largely on the type
of data one starts from and even so there are several avenues that
could be pursued. In this Section we will work towards 
a framework within which to work with
cumulants whenever the data is provided in real space. 

We are interested in probing the statistical
properties of the data set at different scales, in such a way that
the statistics at different scales are independent of 
each other as well as attempting to make them sensitive to local
properties of the map. 

As a first attempt at quantifying the properties of the one point
distribution function one can estimate the cumulants of the pixels
of one data set. Defining the estimators for the moments as above,
one finds that statistical dependence of the data points increases
the variance of each estimator. It can be shown that the largest term
in $1/N$ in the variance of these estimators is modified by
a factor. I.e. we have
\begin{eqnarray}
\langle |{\hat \chi}_r|^2\rangle \simeq
\frac{r!}{N_{eff}}\sigma^{2r}
\label{eq:samdep}
\end{eqnarray}
where 
\begin{eqnarray}
\sigma^2=C(0)
\end{eqnarray}
$N_{eff}$ is defined to be
\begin{eqnarray}
\frac{\sigma^4}{N_{eff}}=\frac{1}{N^2}\sum_{i,j}\langle T_i T_j \rangle^2=
\frac {1}{V^2}\int d^2x \int d^2x' W({\bf x})W({\bf x'})C({\bf x},{\bf x'})
\end{eqnarray}
where V is the sample area and $W$ is the correspnding window function.
This is clearly a problem: the pixels are correlated and therefore
the effective number of independent pixels is  
smaller than the actual number of pixels.

As discussed in the introduction, statistical isotropy,
homogeneity and Gaussianity lead to the statistical independence 
of $\frac{\Delta T}{T}$ for different wavenumbers. It
is also natural to expect that different physical processes will
predominate at different scales. So ultimately one would
like to define a statistics out of the cumulants which are
scale dependent. In the following we present three alternatives
and argue that the last one has the most advantages
\subsection{Filtered cumulants}

We would like to define a linear transformation that
will take the data and filter out everything but the scales of
interest.  We shall
consider the simplest case for the moment: a real space filter
which filters everything but a band of width $\sigma_q$ around
a wave-mode $q$. In Fourier space this corresponds to a top-hat
function
\begin{eqnarray}
{\tilde W}_q(k)= \left\{\begin{array}{ll}
\frac{1} {\pi q\sigma} &\mbox{if $|q-k|<\frac{\sigma}{2}$} \\
0 &\mbox{otherwise} 
\end{array}
\right.
\end{eqnarray}
The corresponding transformation in real space is given by
\begin{eqnarray}
T({\bf x})&=&\int dy^2 W_q(|{\bf x}-{\bf y}|)T({\bf y}) \nonumber \\
W(r)&=&\frac{1}{4\pi\sigma q}
 \{q[J_1(q_+r)+J_1(q_-r)]+\frac{\sigma}{2}[J_1(q_+r)-J_1(q_-r)]\}
\label{eq:win}\end{eqnarray}
with $q_{\pm}=q\pm\frac{\sigma}{2}$
If we want to keep the number of modes constant {\it per} ring,
we have $\sigma \propto q^{-1}$, i.e. $N$ constant.

The procedure can then be the folowing: given a dataset with
a certain number of pixels and geometry, we can identify
the number of independent scales to be probed. For simple
geometries this is straightforward (the size and Nyquist 
frequency give us the range of modes allowed). For each
wavenumber we perform this transformation using window
function eq. \ref{eq:win} to obtain a new set of values
withthe same size as the original. We then define estimators of
the moments using Eq. \ref{eq:mom} and find the cumulants
using eq. \ref{eq:det}. This quantites have some of the
desired properties. Firstly they are stratifying the information
in terms of scales and quantifying the non-Gaussianity in these
bands in wavenumber. Secondly, both the convolution and
the estimators are rotationally invariant so the final 
results are independent of the axis of orientation on
which you are working. Thirdly, the window functions
are chosen to be orthogonal, so one can decouple the
information between rings. So for a given two rings with
different wavenumbers, $k$ and $k'$, the cumulants of
one ring, ${\hat \chi}_r(k)$,
will be independent of the cumulants of the other, ${\hat
\chi}_{r'}(k')$. 

There are strong shortcomings with this approach however.
Firstly it is a highly non-local operation on the data
set. The Fourier transform will is very sensitive to
the global properties and geometry of the data and
it is difficult to separate out what is truly non-Gaussian
and what is due to large scale sampling effects. Of
course this just means that one has to be careful 
when analysing the cumulants of the large wavelength modes
but any form of anisotropy in the sky coverage may
corrupt this analysis down to very small scales.
Secondly it is a very inefficient transformation. For
each wavenumber we produce a set of N points on which we
define the estimators. The fact that the statistics
is defined on these N points is misleading: for a
filter with a wavenumber $k$ one should have approximately
$N_k=f_{sky}(2k+1)$ independent points, much less than N.
This will manifest itself if we introduce the correction into
eq. \ref{eq:samdep} with $C(r)$ replaced by
\begin{eqnarray}
C_k(r)= W^2_k \star C(r)
\end{eqnarray}
where $\star$ indicates convolution. 
\subsection{Cumulants of continuous wavelet transforms}

We are interested in defining a set of linear transformations that
filter information in fourier space but at the same time keep
information about localization. We can see the Fourier representation
as one extreme, where we operate on the whole data-set, so that each
Fourier value is very non-local. The other extreme is
configuration space representation, where the basis vectors are
$\delta$-functions on the pixels.

There is a framework within which one can have basis functions
which are both localized in Fourier space and in real space. These
are called wavelets \cite{wavelets}. In brief the idea is the
following. 
One can expand a function $f$ in terms of a set of basis functions:
\begin{eqnarray}
f({\bf x})=\sum_{l,j}{\tilde f}({\bf x}_l,k_j)\varphi_{l,j}
({\bf x})
\end{eqnarray}
where the index $j$ labels the frequency band one is exploring,
and the label $l$ labels the position band one is probing. 
These functions are compactly or almost compactly supported
in real space, so they evaluate local features. One can construct
these functions from a basic building block, the {\it parent}
function, $\varphi({\bf x})$ through a set of translations and dilations. This
parent function must satisfy an admissibility condition
\begin{eqnarray}
\int\varphi(x)dx=0
\end{eqnarray} A good
example, for our purpose is that of the Maar wavelet in two 
dimensions:
\begin{eqnarray}
\varphi({\bf x})=(2-x^2)e^{-\frac{x^2}{2}}\label{eq:maar1}
\end{eqnarray}
To generate a family of functions with the same features
we define
\begin{eqnarray}
\varphi_{l,j}({\bf x})=\varphi(\frac{{\bf x}-{\bf x}_i}{k_l})
\label{eq:maar2}
\end{eqnarray}
A transform of the data is then
\begin{eqnarray}
\frac{\Delta \tilde T}{T}({\bf x}_l,k_j)=\int d^2x\frac{\Delta T}{T}({\bf x})
\varphi_{l,j}({\bf x})\label{eq:cwt}
\end{eqnarray}
The transform is isotropic and one can estimate, given a mode $k$
that the numebr of points ${\bf x}_i$ one should obtain is
${\tilde N}\simeq \frac{V}{k^2}$, much less than in the previous
section. It is also true that the quasicompact support of these
functions avoid problems with the irregular geometries that one faces
with just a straight Fourier transform. For a given wavenumber
one simply packs the transformed regions in such a way as
to avoid the boundaries.

Although this is a considerable improvement in many respects
to a full Fourier transform there is one problem. As yet
there is no systematic way of defining an orthonormal
set of functions. The prescription described above
will generate a very large number of basis functions
which are all interdependent. This means that, if we
define the cumulants for each band in wavenumber and use
eq. \ref{eq:mom} where we replace the sum over pixels
by the sum over the spatial coefficients of the transform
(the ${\bf x}_l$' in eq. \ref{eq:cwt}) then not only will
cumulants within one ring be correlated, but cumulants
within adjacent rings will be correlated. 

\subsection{Cumulants of discrete wavelet transform}\label{dwt}
It turns out that there is framework within which one
construct an orthonormal set of functions. One can define a
set of functions in one dimension:
\begin{eqnarray}
\varphi_{j,i}=2^{\frac{j}{2}}\varphi(2^j-i)
\end{eqnarray}
where $\varphi(x)$ satisfies the admissability condition in
one dimension. These functions are orthonormal. To complete
the basis one needs an additional function which satisfies
$\int dx\phi(x)=1$ to contain information about the low
frequency modes. Then any function can be expanded in terms of
$\varphi_{i,j}$ and $\phi_i(x)=\phi(x-i)$.  There is
a certain freedom in constructing such a set of functions
but Daubechies has proposed an efficient algorithm for
such a purpose, which we shall outline. To define $\varphi(x)$
and $\phi(x)$ one can solve a set of equations:
\begin{eqnarray}
\phi(x)&=&\sum_ic_i\phi(2x-i) \nonumber \\
\varphi(x)&=&\sum_i(-1)^{1-i}c_{1-i}\phi(2x-i) \nonumber \\
|c_i|^2&=&2-\frac{(2D-1)!}{(D-1)!2^{2-2D}}\int_0^i\sin^{2D-1}xdx
\end{eqnarray}
The integer D dictates the size of the compact support of the
wavelet and also indicates the regularity of the wavelet (i.e.
the number of coefficinets which are zero in the Taylor
expansion of each wavelet).

These functions have the properties that we have been looking for.
They are local (they have compact support in real space) so one
becomes insensitive to global features of the sample. They
are orthonormal so one is separating information between different
modes. This transform is efficient and will not give us redundant
information; from $N$ pixels we will obtain $N$ coefficients.
As yet, no higher dimensional analogue of this transfrom has
been developed with the same useful properties. We can, however,
construct 2-dimensional wavelets using the tensor products
of one dimensional wavelets i.e.
\begin{eqnarray}
\phi_{i_1,i_2}(x_1,x_2)&=&
\phi_{i_1}(x_1)\phi_{i_2}(x_2)\nonumber \\
\varphi_{i_1,i_2;j_1,j_2}(x_1,x_2)&=&
\varphi_{i_1;j_1}(x_1)\varphi_{i_2;j_2}(x_2)\nonumber \\
\xi^1_{i_1,i_2;j_1}(x_1,x_2)&=&
\varphi_{i_1;j_1}(x_1)\phi_{i_2}(x_2)\nonumber \\
\xi^2_{i_1,i_2;j_2}(x_1,x_2)&=&
\phi_{i_1}(x_1)\varphi_{i_2;j_2}(x_2)
\end{eqnarray}
In using this approach we can now define rotationally and
translationally invariant estimators for the moments.
We define
\begin{eqnarray}
{\hat \mu}^r(k)=\sum_{i_1,i_2}\sum_{j_1,j_2}
\left(\frac{\Delta T}{T}\right)^r_{i_1,i_2;j_1,j_2} \ \ \ \mbox{where 
$2^{-j_1}+2^{-j_2}=k$}
\end{eqnarray}
and then use eq. \ref{eq:det} to define the cumulants. 

\section{Fourier Space Statistics}\label{fouriers}
For the sake of completeness we shall now  discuss a possible
application of cumulants to interferometric measurements, i.e.
data in Fourier space \cite{catpeople}. Fourier mode
cumulants allow an alternative formulation of the ring spectra
as defined in \cite{fermag}. The construction in \cite{fermag}
is based on dividing Fourier space into rings (scales) with $\Delta k=1$.
In each ring live a total number $N(k)=2kf_{\rm sky}$ of modes $\{a({\bf k})\}$
(counting the real and imaginary parts separately)
which are all independent for any Gaussian theory. In this formula $f_{\rm sky}$ is
the fraction of sky covered by the experiment.
The statistical independence of these modes is a mere
implication of the orthogonality of the Fourier functions,
and is in contrast with a real space formulation, where pixels
are generally correlated. Statistical independence facilitates
computing the effective number of independent modes, avoiding the
trouble described in Section~\ref{real}. In the same spirit, 
and following \cite{hobsmag} one can define an uncorrelated mesh
of independent Fourier modes. The
effective number of independent modes {\em within} a mesh cell
centred on $\bf{k}_i$ is given by
\[
\widetilde{N}_{\rm cell}(\bf{k}_i)=
1/ \int_{\rm cell}{d^2\bf{k}\over A_{\rm cell}}
\{{\rm cor}[a_s(\bf{k}_i),a_s^*(\bf{k})]\}^2,
\]
where $A_{\rm cell}$ is the area of the mesh cell. Clearly, 
$\widetilde{N}_{\rm cell}(\bf{k})$ is always greater than unity. 
One can therefore, on average, avoid the 
loss of non-redundant information by computing the average
density of independent modes around a given mode at $\bf{k}=\bf{k}_i$
\[
\rho(\bf{k}_i)=1/ \int d^2\bf{k}
\{{\rm cor}[a_s(\bf{k}_i),a_s^*(\bf{k})]\}^2,
\]
and defining the mesh size as $k_0=1/{\sqrt{\rho}}$. In this way
estimators of statistics derived from large regions of the $\bf{k}$-plane
have the same variance whether one uses the mesh
or the continuum of modes in its calculation. The size of the
mesh cell is $k_0\approx 2\pi/L$ for a square patch with size $L$,
whereas for the Gaussian window of an interferometer it is
$k_0={\sqrt {2\pi}}/\sigma_w$ where $\sigma_w$ is the variance of
the Gaussian (so that its FWHM is $\theta_w\approx 2.3 \sigma_w$).

Let us now look at the real and imaginary parts of the mesh modes
living in each ring with $\Delta k=1$. We may then compute
the cumulants of this sample $\chi_r(k)=\chi\{\Re a({\bf k}),\Im a({\bf k})\}$.
thereby producing a two index spectrum $\chi_r(k)$.
This spectrum includes the power spectrum ($r=2$) but also complements it 
with information on how the power is distributed in each ring,
encoded in the $r>2$ components. In the language of \cite{fermag}
they complement a $C_\ell$ spectrum with a set of $B^\ell_m$ 
telling us how the power is distributed between all the modes 
in the ring (in direction and phase). Hence the $r$ dimension of
the $\chi_r(k)$ spectrum may be seen as a ring spectrum, as the one
proposed in \cite{fermag}.

The cumulants $\chi_r(k)$ share some of the properties given before for
real fields. They can be estimated from ${\hat \chi}_r(k)$ as before.
Their covariance matrix of the estimators for a Gaussian process takes the form:
\begin{equation}
\langle {\hat\chi}_r(k) {\hat\chi}_{r'}(k')\rangle =\delta_{kk'}{\left(
\frac{r!}{N(k)}\mu_2^{r}
\delta_{rr'}+{\cal O}(\frac{1}{N^2})\right)}
\end{equation}
The result for the power spectrum ($r=2$) is well known
\begin{equation}
\langle {\hat\chi}_2(k) {\chi}_{2}(k')\rangle =\delta_{kk'}
\frac{2}{N(k)}
\end{equation}
Up to $1/N^2$ all that changes is the coefficient $2$ in this formula, as
we go to higher order cumulants. The new coefficient is $r!$ so the
variance of higher order cumulants increases very quickly. By requiring
that this variance be smaller than a given value we impose a data-reduction
criterion, which ensures that we will end up with a number of $\chi_r(k)$
smaller than $N(k)$ for each $k$.

Cumulants have advantages over the variables $\theta$ defined in
\cite{fermag}. To begin with they are additive. This enables computing
angular spectra in bands, rather than rings, thus profiting from an
enlarged number of modes. This technique leads to what is essentially
direct filtering in Fourier space. It also allows the definition of
something like a $B_m$ rather than a $B^\ell_m$, which is in a way
a more orthogonal description to the power spectrum $C_\ell$.
On the other hand cumulants have a disadvantage over
the variables $\theta$ defined in \cite{fermag}: their distribution
for a Gaussian is not simple. All we have computed for them is a covariance
matrix, which is clearly not the end of the story, 
because their distribution is
not Gaussian. The variables $\theta$ on the other hand are simply uniformly
distributed. Also the variables $\theta$ can never exceed in number 
the initial number of modes, whereas a cosmic variance criterion
must be introduced for cumulants in order to introduce a truncation.

Unfortunately the ring cumulants $\chi_r(k)$, although invariant under
rotations, are not invariant under translations.
What is even worse, their average square variation under
translations is always comparable to their variance for a Gaussian
process, even for large $N(k)$. For simplicity we shall illustrate this with
the moments ${\hat\mu}_r(k)$. Their variance can be easily shown to be 
of order:
\begin{equation}\label{varmu}
\sigma^2({\hat\mu}_r(k))=\frac{\mu_{2r}-\mu_r^2}{N(k)}
\end{equation}
Under translations the real and imaginary parts of each mode get rotated
by and angle equal to $\delta\phi={\bf k}\cdot{\bf t}$ mod $2\pi$. For 
a uniformly distributed translation this induces an average 
change in ${\hat\mu}_r(k)$ of the order of $\delta {\hat\mu}_r(k)/\delta\phi$.
This is of course zero. However the mean square of the change in
${\hat\mu}_r(k)$ is
\begin{equation}
{\langle {\left({\delta{\hat\mu}_r(k)\over \delta\phi}
\right)}^2\rangle}=\frac{2r^2(\mu_{2r-2}\mu_2
-\mu_r^2)}{N(k)}
\end{equation}
comparable to (\ref{varmu}). We may regard this as a problem, or not. 
Measurements in Fourier space usually use small fields.
Small fields break translational invariance by the mere existence of
a window. 

\subsection{Fourier space filtering}
One can take advantage of the fact that measurements are in Fourier
space to perform direct filtering in Fourier space. As we have emphazised
there are situations where the non-Gaussian signal is contaminated by a 
Gaussian signal. It may further happen that the signal non-Gaussianity is better 
isolated in Fourier space, that is, non-Gaussianity dominates in some scales
and is dominated in others, with a clear separation of these two regimes.
If data is in Fourier space in the first space then all we should do is
find a window $W(k)$ defining the band where non-Gaussianity is the purest.

Again the cumulants additivity will help to quantify the effect this has 
on the cumulants. Let us compute cumulants of a sample made up of 
the real and imaginary parts of all the modes inside a given band, weighed by
a window $W(k)$. Such a quantity could be related to the ring estimators 
${\hat \chi}_r(k)$ by
\begin{equation}
 {\hat\chi}_r(W)=\int d k\,  {N(k)W(k)\over N(W)}{\hat\chi}_r(k)
\end{equation}
where $N(W)$ is the total number of modes in the band. It would therefore
average to 
\begin{equation}
 \chi_r(W)= \int d k\,  {N(k)W(k)\over N(W)}{\chi}_r(k)
\end{equation}
If the signal is purely non-Gaussian and it involves a non-zero average
cumulant which does not change sign over the band, then extending the sample
over the whole band simply accumulates non-Gaussian signal in the cumulant.
One is simply integrating a function which does not change sign over a 
domain thereby making the result more different  than zero.

Furthermore the confusion with a Gaussian process also decreases
because the error bars around zero for a Gaussian process also get
smaller if one uses the whole band as a sample. It is easy to see that
for a Gaussian random field
\begin{equation}
  \sigma^2({\hat\chi}_r(W))={r!\mu_2^r\over N(W)}
\end{equation}
For the very simple reason that we have more modes inside the band than inside
each ring the error bar around zero for a Gaussian is much smaller for
a band than for any ring. 

For these two reason it makes sense computing band cumulants rather than
ring cumulants.
In work in progress we make use of this technique in the search of cosmic
strings by interferometers \cite{lewin}.

\subsection{Connection with real space statistics}
Finally we should add that 
the ring moments $\mu_r(k)$ and cumulants  $\chi_r(k)$ allow a quick
connection to some simple real space statistics based 
on histograms of temperature derivatives.
If the non-additive moments are used these can be related to
the Fourier transform of the $n$-point correlation function
using (\ref{fourier}):
\begin{equation}
  \mu_r(\partial_i \cdots\partial_j\frac{\Delta T}{T})
  =\int \frac{dk_2}{(2\pi)^{1/2}}\cdots \frac{d{\bf k}_n}{2\pi}
     i^r k_i\cdots k_j C^r(k_2, \cdots,{\bf k}_r)
\end{equation}
The $\mu_r$ of temperature derivatives can therefore be connected with the
the $n$-point correlation function $C^n(k_2, \cdots,{\bf k}_n)$ 
but not with the ring moments $\mu_r(k)$. 
These $\mu_r$ are promising as they integrate over the redundant
degrees of freedom in the $n$-point function $C^n(k_2, \cdots,{\bf k}_n)$.

Using the cumulants on the other hand one has
\begin{equation}
  \chi_r(\frac{\Delta T}{T})=\int dk\, 2\pi k \chi_r(k)
\end{equation}
or for the temperature derivatives
\begin{equation}
  \chi_r(\partial_i \cdots\partial_j\frac{\Delta T}{T})=
  \int dk\, i^n k_i\cdots k_j 2\pi k \chi_r(k)
\end{equation}
We see that the $\chi_r$ of the temperature derivatives
consist of integrals of ring histograms $\chi_r(k)$ subject to different weighting
powers of $k$. These powers of $k$ can be seen as a Fourier space filter.
In fact, what the Fourier space filtering for cumulants, which advocated
above, is doing is generalizing these statistics to filters other than
power laws in $k$.

This immediately suggests a way to convert Fourier space filters
into real space statistics. For filters other than power laws
one obtains  linear operations on the temperature
maps other than the derivatives, but the practical procedure
is essentially the same.
Let $W(k)$ define the ring where non-Gaussianity is the purest.
We may then define the optimized  statistic
\begin{equation}
  \chi^{opt}_r=\int dk\, 2\pi k W(k)\chi_r(k)
\end{equation}
This filter may then be inverted into a real space statistic by means of 
\begin{equation}
  \chi^{opt}_r=\chi_r(\frac{\Delta T}{T}\star W(x))
\end{equation}
where $W(x)$ is the Fourier transform of the window $W(k)$.

\section{An Application}
To illustrate the use of the method in real space we shall construct
a simple example: a Gaussian CMB signal superposed onto a Poisson
distribution of point sources. As mentioned above, for certain
frequencies, and for high resolution, the power spectrum of the
Gaussian signal is exponentially damped due to the finite thickness
of the surface of last scatter. The point source distribution, however
will be white noise and so may dominate on small scales. We then have
a signal dominated on large scales by the Gaussian source which
may obscure the small scale non-Gaussianity. This is an ideal scenario
in which to apply our technique.
\begin{figure}
\centerline{\psfig{file=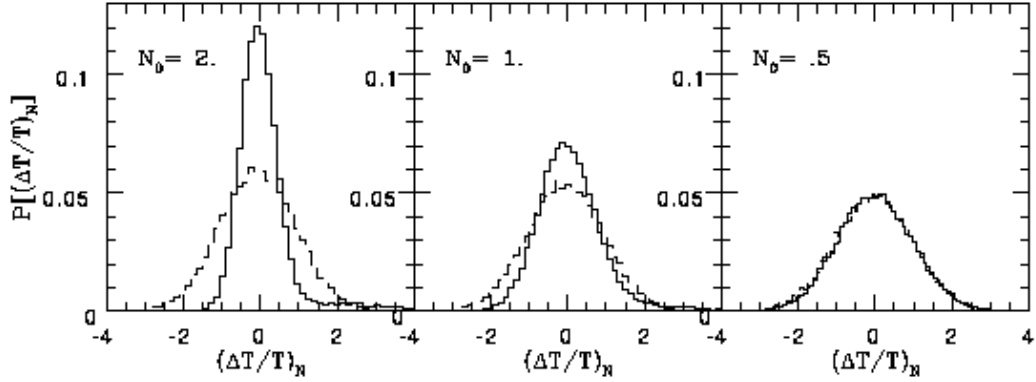,width=6.in}}
\caption{Histograms of pixels for three relative amplitudes
($N_0$). In each case the temperature distributions are
normalized to unit variance. 
}\label{his}
\end{figure}

\begin{figure}
\centerline{\psfig{file=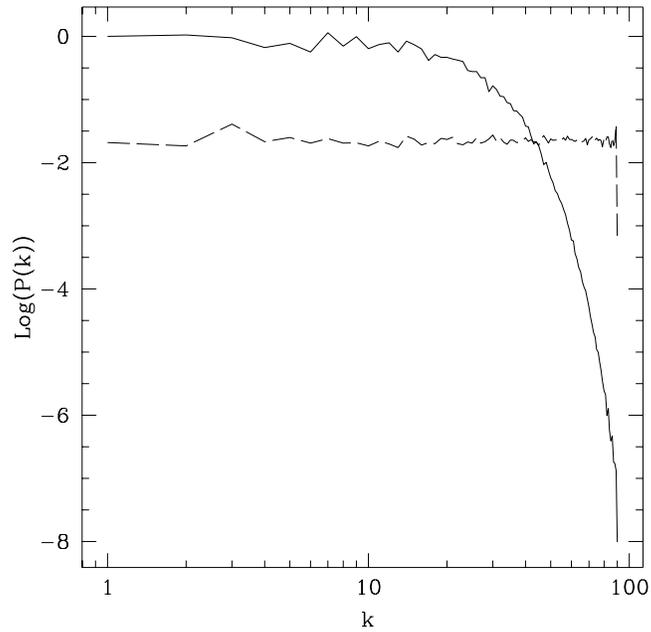,width=3.5in}}
\caption{The power spectra of both the Gaussian (solid line)
and non-Gaussian (dashed line) signal. 
}\label{ps1}
\end{figure}

\begin{figure}
\centerline{\psfig{file=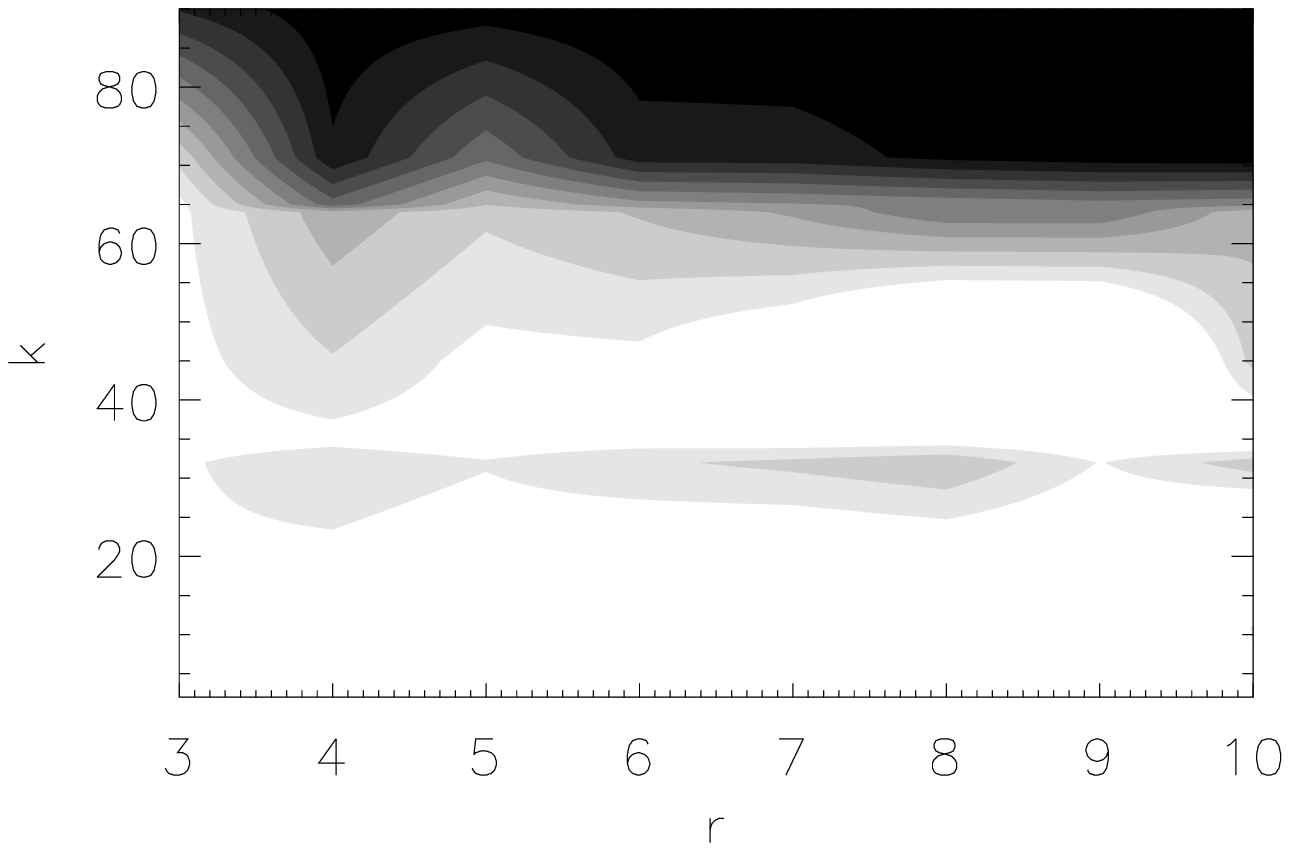,width=3.5in},\psfig{file=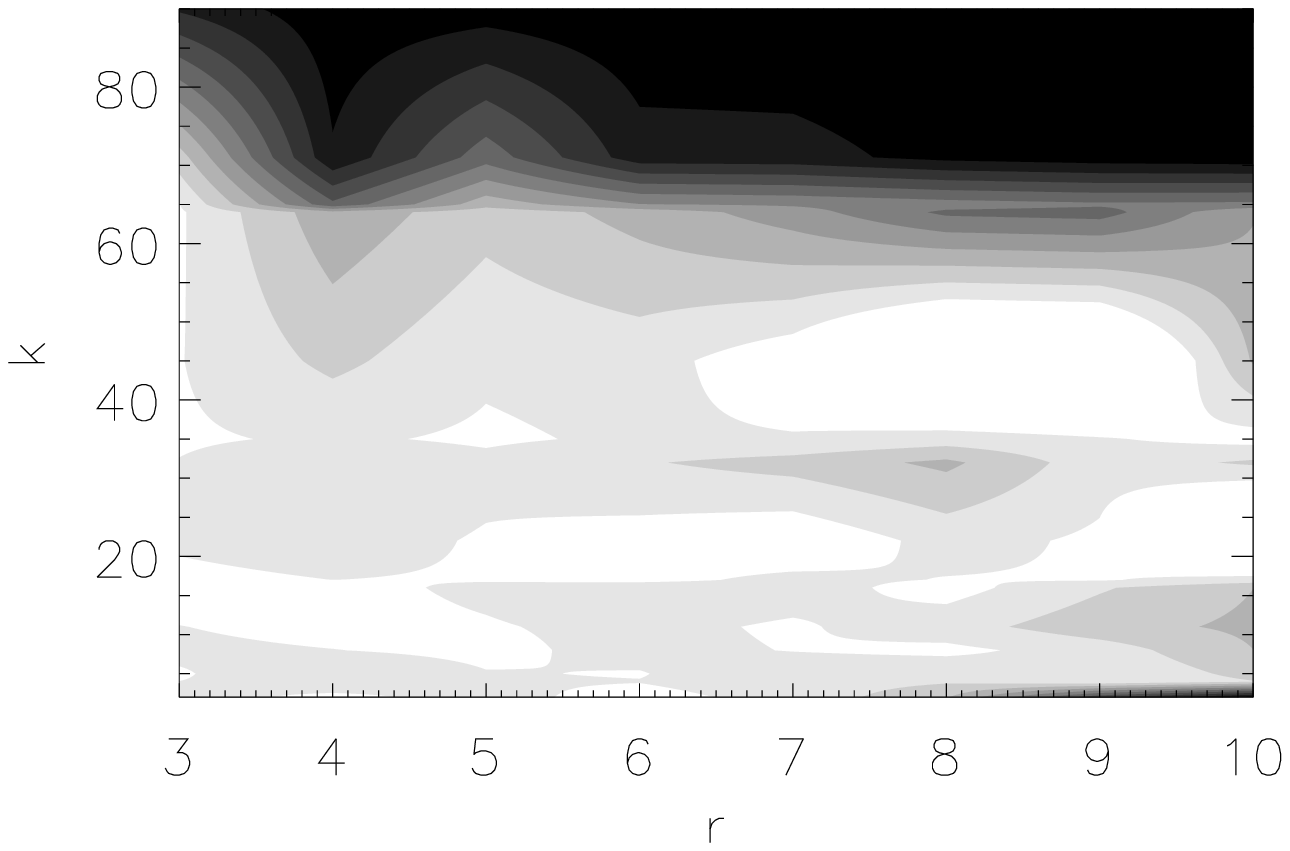,width=3.5in}}
\caption{The excess variance of $\chi_r(k)$, $E_r(k)$ (left panel) and
${\bar \chi}_r(k)$, $E_r^N(k)$ (right panel). The contour lines are at
.1, .2 $\cdots$ 1. , from black to white.
}\label{exc}
\end{figure}

Considerable work has been done in finding the statistical
properties of a field generated by a set of point sources
\cite{scheuer,mandolesi} which has led to the widely used
``$P(D)$'' approach. We shall use the basic ingredients 
described in this work to construct the non-Gaussian source.
We shall generate a Poisson distribution of sources in the
sky, in which the number of sources with intensity $S$ per steradian
is given by a simple fit
\begin{eqnarray}
N(S)=\left\{ \begin{array}{ll} 
0 & \mbox{$S<S_0$} \\
\kappa S^{-\beta} & \mbox{otherwise}
\end{array} \right.
\end{eqnarray}

For purpose of illustration we shall use $\beta= 1.5$.
(for our purposes $k$ and $S_0$ will mostly affect the 
overall normalization). In \cite{scheuer,mandolesi}
an expression for the probability distribution function
of the fluctuations was derived as a function of these
parameters. However we are interested in the additional
complication of superposing a Gaussian signal. The signal
we shall use has a power spectrum 
\begin{eqnarray}
C_G(k)=A\exp(-(k/k_d)^2)
\end{eqnarray}
Therefore the full signal is given by
\begin{eqnarray}
\frac{\Delta T}{T}=\frac{\Delta T}{T}_{ps}+\frac{\Delta T}{T}_{G}
\label{nge}
\end{eqnarray}
where $ps$ ($G$) label the point source (Gaussian) components.
We choose to fix $A$ and $k_d$ from
\begin{eqnarray}
N^2_0&=&\frac{<\left(\frac{\Delta T}{T}\right)_{ps}^2>}
{<\left(\frac{\Delta T}{T}\right)_{G}^2>} \nonumber \\
N^2_{\infty}&=&\lim_{k\rightarrow 0}{{C_{ps}(k)}\over{C_G(k)}}
\end{eqnarray}

By varying $N_0$ and $N_{\infty}$ we can enhance or supress the
non-Gaussian signal. In Fig.~\ref{his} we can see how the shape
of the pixel distribution changes as we change $N_0$ ($N_{\infty}
=.2$ is kept fixed). The solid
line (a histogram of pixels generated as in Eq.~\ref{nge}) gradually
merges with the dashed line (a histogram of pixels of a Gaussian
realization with the same power spectrum). As argued above,
the large scale Gaussian fluctuations are dominating the small
scale behaviour of the non-Gaussian signal, and it is necessary 
to find the region where this is possible.
We choose to explore with the configuration of maximum confusion,
(illustrated in the right hand panel). The parameters are then
$N_0=0.5$ and $N_{\infty}=0.2$ and we show a comparison of
the power spectra  in Fig.~\ref{ps1}. Given the way we normalize the
signal we have found that the statistics
we are analysing in this section are essentially insensitive
to different values of $\beta$.

\begin{figure}
\centerline{\psfig{file=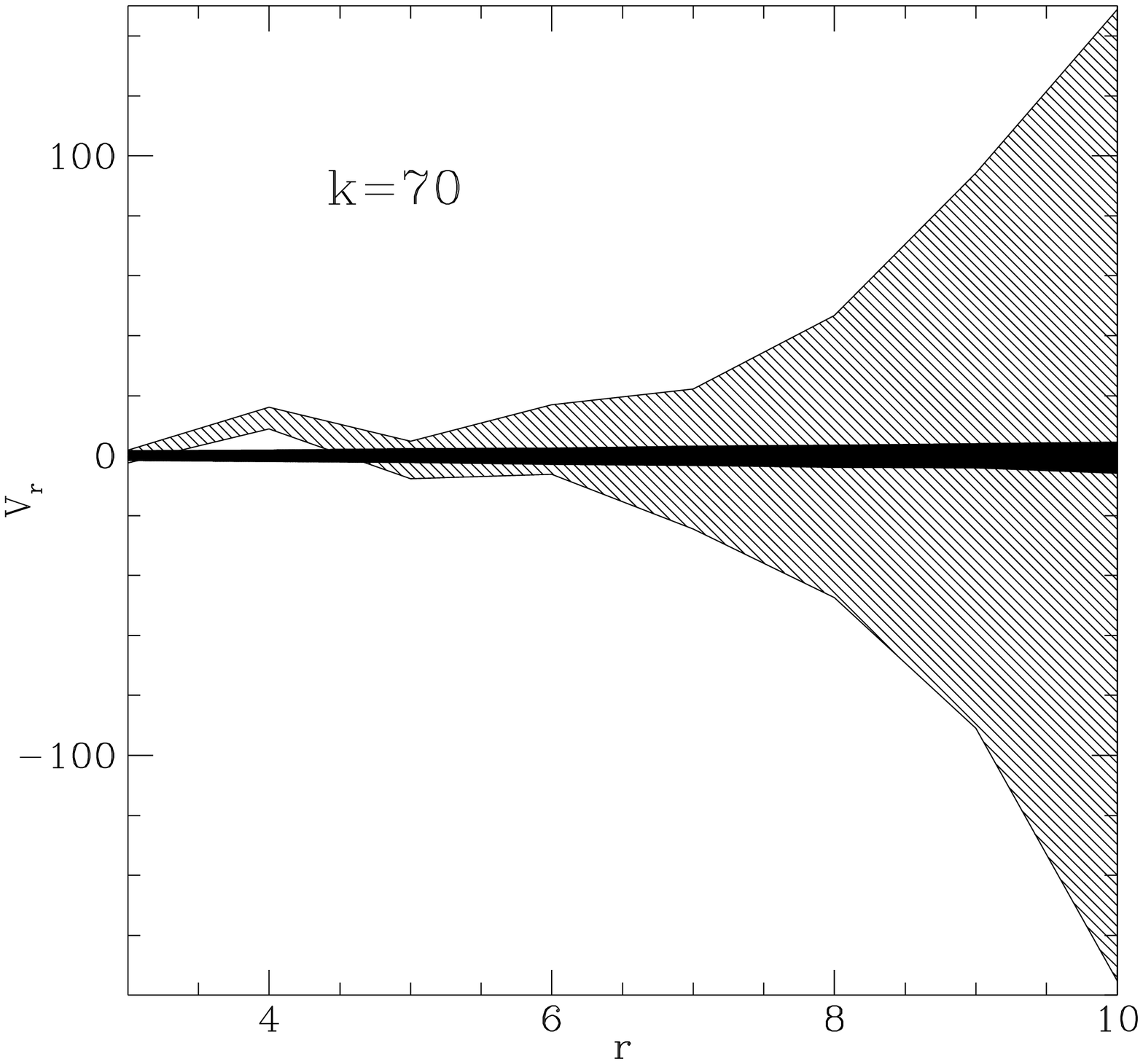,width=3.5in},\psfig{file=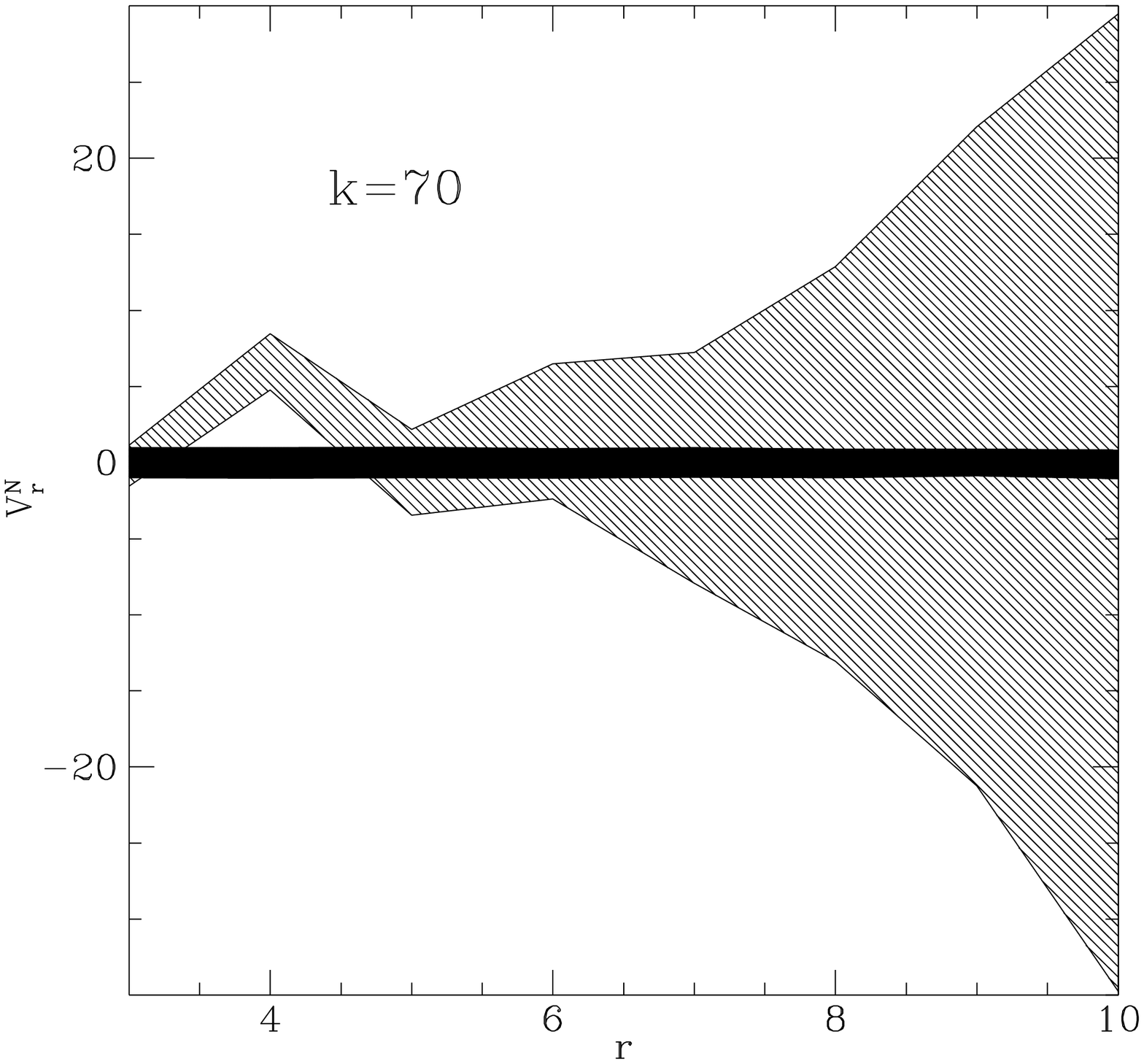,width=3.5in}}
\caption{One $\sigma$ regions for $V_r$ (left panel)
and $V^N_r$ (right panel). The solid band
corresponds to the Gaussian maps and the hatched band to the
non-Gaussian maps.
}\label{var}
\end{figure}

\begin{figure}
\centerline{\psfig{file=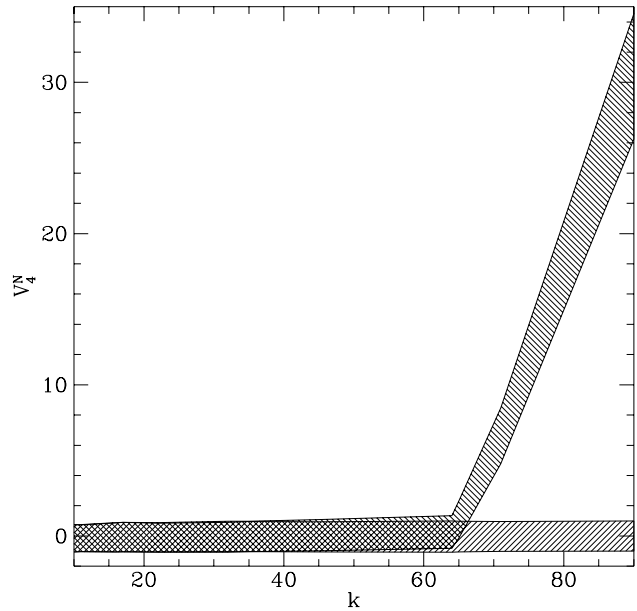,width=3.5in}}
\caption{
One $\sigma$ regions for 
$V_4^N$
the kurtosis (suitably
normalized). The horizontal, hatched region corresponds to the
Gaussian.
}\label{ps5}
\end{figure}

To get a detailed understanding of the statistics we
generate an ensemble of 10000 maps (of $128^2$ pixels)
using Eq.~\ref{nge} and the same number of Gaussian realizations
with the same power spectrum. A natural thing to look at is the
variance of both $\chi_r(k)$ and ${\bar \chi}_r(k)$; in fact
it is instructive to plot the excess variance of the non-Gaussian
distribution with regards to the Gaussian. One can define
\begin{eqnarray}
E_r(k)&=&\frac{\sigma_G[\chi_r(k)]}{\sigma_{NG}[\chi_r(k)]}\nonumber \\ 
E^N_r(k)&=&\frac{\sigma_G[{\bar \chi}_r(k)]}
{\sigma_{NG}[{\bar \chi}_r(k)]}
\end{eqnarray}
and in Fig~\ref{exc} we present a contour plot for both these
quantities. Clearly for large wavenumbers, (for $k>65$),
 the non-Gaussian
distribution has a large excess variance as compared to the Gaussian
one ($E_r$ becomes less than $0.2$ very rapidly).
 For small wavenumbers there is confusion. The fact that we can
pick up such a large difference is due to the fact that we are
looking at the scales where the non-Gaussian signal dominates {\it
and} we are using a set of statistics which preserves local
information.

If we now focus on a band at high wavenumbers we can see the 
difference between the two distributions in more detail. It is useful
to define
\begin{eqnarray}
V_r&=&\chi_r(k)\times \sqrt{\frac{N_{eff}}{\sigma^{2r}r!}}\nonumber \\
V^N_r&=&{\bar \chi}_r(k)\times \sqrt{\frac{N_{eff}}{r!}}
\end{eqnarray}
Note that, for a Gaussian distribution, the
 variance of $V_r$ should be proportional to $\sigma^r$ while
the variance of $V^N_r$ is one. In
Fig.~\ref{var} we plot the $68\%$ confidence regions for
these two quantities for $k=70$. The excess
variance, again is manifest and there is a strong signal
of non-Gaussianity.

From Fig.~\ref{var} we can see that there is a  value of
$r$ for which the two bands do not overlap. It corresponds
to the (appropriately normalized) kurtosis of the distribution.
If we concentrate on $V_4^N$ as a function of scale we see that,
as we expect, for low wavenumbers the two distributions 
are indistinguishable while for wavenumbers larger than
$k=65$ there is a large discrepancy (see Fig.5). The same 
cannot be said about $V_3^N$, the skewness. 

One may learn important lessons from this example. Introducing a scale
into our statistics was clearly a good idea. Making a pure histogram of
pixels was found hopeless, but filtering the data into a hierarchy
of scales allowed the recognition of the point sources, by inspecting
the appropriate band of scales. In that band two types of fingerprints were
found for point sources. For the 
kurtosis there is clearly a positive
average, with no overlapping cosmic variance error bars with a Gaussian
process. For higer order cumulants the signal is more subtle. Although
these cumulants still average to zero they show an abnormally large spread.
Therefore in most realizations one would find a value for the cumulant
well outside the Gaussian cosmic variance error bar, even though
averaging over realization still leaves a zero cumulant.
This is an example of a situation where the variance errorbar is more
important than the average quantity.
These two signatures are clear and strong indications of the
point source non-Gaussianity. 

Another important lesson is the advantage of recognizing local rather
than global features. The signatures found above do not get
complicated by adding more and more point sources. They are essentially
dependent only on the non-Gaussian features of the individual structures.
This is an advantage over the treatment of point sources given 
in \cite{fermag}, which was really only simply when there was 
a single point source inside the field (a situation common in the very
small field context analyzed in that paper). The non-Gaussian spectra 
defined in \cite{fermag} recognize global rather than local shapes.
Hence if there were many point sources in the field they would
recognize the angles  between the lines connecting the various point
sources, and the lenghts of all the segments, 
rather than the point sources themselves. For instance if
there were three point sources in the field the formalism would react
to the shape and size of the triangle depicted. This is naturally 
a complete mess for a Poisson process, even though the individual
objects are very simple. The formalism used in this section, on the
other hand, always recognizes individual structures. This is achieved
both by the use of the wavelet transform, and the use of additive
cumulants, and is a desirable  feature whenever the trees are simple but
the forest is complex.

\section{Discussion}
In this paper we have developed a new technique for
quantifying non-Gaussianity using cumulants. Although in 
Section~\ref{fouriers} we have
discussed the possibility of using these quantities on
interferometric data, we have focused on applying it to
real space data. In this setting we have defined a
set of statistics which are scale dependent, rotationally
invariant and which are computationally easy to evaluate.
In the limit of large data sets they are statistically
independent for a Gaussian random field. One has the option
of working with power spectrum independent quantities
or additive quantities depending on preferences.
This work is the first step in defining a useful set of statistics
for analysing the future large data sets from ground and space
based CMB experiments. The next steps are obvious and we
shall discuss the prospects of each. 

We have taken the large$-N$ limit of our data sets and this
has allowed us to define simple estimators and find the
simple structure of the covariance matrix in Eq.~\ref{covcu}. 
Although this is the case of satellite
experiments, ground based experiments in the near future
may not satisfy this condition. It becomes necessary then to
analyse the case of moderate $N$ and a number of problems arise.
To begin with the estimators defined in Eq.~\ref{eq:mom} are
biased and not centred. This means that the ${\hat \chi}_r$s
won't have zero expectation values for a Gaussian process. 
However, as mentioned above
there is a standard procedure for dealing with this using
$k-$statistics, i.e. defining statistics which have the correct
expectation value. The problem then arises that the covariance
matrix loses its simple form. In particular the off-diagonal
terms become non-negligible. In the same way that
the dependence in $r$ and $N$ gave us a criteria for truncating the
number of moments to calculate, one can now impose the condition
of effective diagonalization. I.e by defining a how large
the off-diagonal terms are allowed to be relative to the diagonal
one again obtains a constraint on $r$ given $N$. An alternative,
slightly more convoluted approach is to construct linear
combinations of the ${\hat \chi}_r$ so that the covariance matrix
becomes diagonal. The interpretation of these new quantities
is less clear.

The formalism we have developed is applicable in the small
angle limit, when the sky can be approximated by a plane.
Given the existence of an all sky data set (from the COBE
satellite) and the expected results from the planned satellite
missions, it is necessary to extend this construction to the
spherical spaces. Although there has been some progress
in developing wavelet techniques on arbitrary surfaces,
work on fast discrete wavelet transforms in such setting is still in its
infancy. There have been some proposals \cite{sweldens}
and the current rate of progress is such that efficient
algorithms will be available in the near future.

We have considered a simple example with which to illustrate
our technique. In considering a non-Gaussian signal from a
distribution of point sources we have made contact with
the $P(D)$ approach of \cite{scheuer,mandolesi}. Indeed,
as mentioned in Section~\ref{cuhi} cumulants are the algebraic
way of characterizing a distribution. By looking at the
$P(D)$ one is essentially looking at a histogram of temperature
fluctuations and as we have argued, calculating the cumulants is
the natural next step.

We have restricted ourselves to the two dimensional fields of
temperature anisotropies. However the estimators we have defined here
can be defined in any dimensions. Such situations have been
explored in \cite{lhizhi} where the statistical properties
of Ly$\alpha$ clouds (one dimensional data sets) were studied
in some detail. One could also envisage performing the same
sort of analysis on three dimensional fields, such as the
distribution of matter in the universe \cite{lascampanas}. Indeed with the
planned large scale surveys of galaxies it should be
possible to characterize the distribution of density perturbation
with unprecedented precision. 

\section*{Acknowledgements}
We thank S. Hanany, M. Hobson, A. Jaffe and J. Levin for useful discussions.
J.M. thanks MRAO-Cambridge for use of computer facilities while 
this paper was being prepared.
P.F. was supported by  the Center for Particle Astrophysics, a NSF Science and
Technology Center at UC Berkeley, under Cooperative
Agreement No. AST 9120005. J.M. was supported by a Royal
Society University Research Fellowship.

\pagebreak
\pagestyle{empty}

\end{document}